\documentclass[12pt,a4paper]{article}

\usepackage{amsmath}
\usepackage{graphicx}
\usepackage[utf8]{inputenc}
\usepackage{booktabs}
\usepackage{url}

\begin{document}
% =====================================================================
% Title, Abstract, and Keywords
% =====================================================================

\title{Bounded by Risk, Not Capability: Quantifying AI Occupational Substitution Rates via a Tech-Risk Dual-Factor Model}

\author{
  Shuyao Gao\thanks{Doctoral Student, aSSIST University, Seoul, South Korea. Email: \texttt{gaoshuyao.mk@gmail.com}. Code and data: \protect\url{https://github.com/ShuyaoGao/bounded-risk-oai}.}
  \and
  Minghao Huang\thanks{Professor, aSSIST University, Seoul, South Korea. Corresponding author. Email: \texttt{mhuang@assist.ac.kr}.}
}
\date{\today}

\maketitle

\begin{abstract}
The deployment of Large Language Models (LLMs) has reignited concerns about technological unemployment. Existing task-based evaluations measure theoretical ``exposure'' to AI capabilities but ignore the frictions of real-world commercial adoption: liability, compliance, and physical safety. We argue that occupations are not eradicated instantaneously; they are gradually encroached upon, action by action. We introduce a Tech-Risk Dual-Factor Model to re-evaluate this process. We deconstruct 923 occupations into 2,087 Detailed Work Activities (DWAs) and use a multi-agent LLM ensemble to score both technical feasibility and business risk. A variance-based Human-in-the-Loop (HITL) validation with an expert panel reveals a cognitive gap: isolated algorithmic probabilities fail to capture the ``institutional premium'' that experts bring to bear under professional liability. Applying a strictly algorithmic baseline via mathematical bottleneck aggregation, we compute Relative Occupational Automation Indices ($OAI$) for the U.S. labor market. Our findings challenge the Routine-Biased Technological Change (RBTC) hypothesis. Non-routine cognitive roles dependent on symbolic manipulation, such as Data Scientists, face unprecedented exposure ($OAI \approx 0.70$). Unstructured physical trades and high-stakes caretaking roles, by contrast, exhibit absolute resilience, quantifying a ``Cognitive Risk Asymmetry.'' We hypothesize the emergence of a ``Compliance Premium,'' under which wage resilience becomes increasingly tied to risk-absorption capacity. We frame these findings as a cross-sectional diagnostic of systemic vulnerability, and as a foundation for subsequent Computable General Equilibrium (CGE) econometric modeling that incorporates dynamic wage elasticity and structural labor reallocation.
\end{abstract}

\textbf{Keywords:} Artificial Intelligence, Labor Economics, Technological Forecasting, Task-Based Approach, Human-in-the-Loop, Risk Aversion.

% =====================================================================
% Chapter 1: Introduction
% =====================================================================
\section{Introduction}

The advent of highly capable Large Language Models (LLMs) and autonomous generative agents has destabilized the contemporary understanding of technological substitution in the labor market. Macro-innovations such as steam power, electricity, and early computing were classified historically as General Purpose Technologies (GPTs) \cite{bresnahan1995general} that reshaped macroeconomic growth trajectories \cite{aghion2017artificial, korinek2018artificial}. The current explosion of generative AI exhibits the hallmarks of a new GPT. Unlike previous waves of mechanization and digitalization, which automated physical labor and routine clerical work \cite{Autor2003}, the current paradigm shows unprecedented proficiency in symbolic manipulation, semantic generation, and non-routine cognitive processing \cite{Noy2023, floridi2020gpt}.

This rapid capability overhang has fuelled a pervasive macroeconomic anxiety, often framed as the imminent threat of sudden occupational obsolescence. The panic has been catalysed by high-profile industry forecasts projecting the disruption of hundreds of millions of global jobs \cite{Briggs2023}. We argue that viewing AI integration through the lens of sudden occupational extinction is analytically flawed. Occupations are not monolithic entities; they are complex bundles of heterogeneous actions. As technological capabilities expand, occupations are not instantly replaced, but gradually encroached upon, action by action, in a phenomenon we call \textit{Task Encroachment}.

Recent empirical work has measured the ``latent exposure'' of the labor market to Generative AI and reached a broad consensus: non-routine cognitive tasks, long insulated under the Routine-Biased Technological Change (RBTC) paradigm, are now the primary frontier of automation \cite{Acemoglu2024, Brynjolfsson2023, Eloundou2023, Felten2023, Noy2023}. This nascent ``Exposure School'' typically quantifies the percentage of task workflows susceptible to LLM integration but stops short of the organizational and regulatory frictions that govern actual deployment. These models therefore conflate \textit{technical feasibility} with \textit{commercial viability}.

Acemoglu \cite{Acemoglu2024} warns that projecting productivity gains straight from task exposure overlooks the micro-frictions and ``hard-to-learn'' tasks that govern real-world integration. A generative model may have the technical capability to draft a binding legal contract or a diagnostic medical script, but adopting it introduces severe legal, ethical, and physical liabilities. In high-stakes environments, current statistical-fitting AI functions as a ``stochastic parrot'' rather than a causal reasoner \cite{Bender2021}, and is therefore prone to unpredictable hallucinations in long-tail edge cases \cite{Ji2023, bommasani2021opportunities}. This probabilistic behaviour collides with the absolute inflexibility of human legal accountability.

This study addresses this critical gap by introducing the \textbf{Tech-Risk Dual-Factor Model}. We hypothesize that true occupational replaceability is governed not solely by the capability of the algorithm, but by the risk tolerance of the commercial environment. To test this, building upon the foundational economic framework that technological displacement occurs at the task level rather than the macro-occupational aggregate \cite{Acemoglu2018}, we dismantle the occupational taxonomy into its most atomic units: Detailed Work Activities (DWAs). Utilizing a multi-model AI ensemble validated by a rigorously stratified, multi-national Human-in-the-Loop (HITL) protocol involving 31 cross-disciplinary experts, we assess 2,087 DWAs across both their technical susceptibility to AI and their inherent business risk.

This methodology allows us to unearth the cognitive divergences between algorithmic probability and human loss aversion, ultimately generating a highly calibrated Occupational Automation Index ($OAI$) for the entire labor market. By mapping the anatomy of task encroachment, this paper aims to transition the discourse from theoretical AI exposure to practical, risk-adjusted labor market restructuring, providing actionable insights for policymakers, educators, and organizational leaders.
% =====================================================================
% Chapter 2: Literature Review
% =====================================================================
\section{Literature Review}

\subsection{The Task-Based Approach and Routine-Biased Technological Change (RBTC)}
The theoretical foundation for analyzing technological impacts on employment is the Task-Based Approach, formalized by Autor, Levy, and Murnane (2003) and later expanded by Acemoglu and Autor (2011). This framework treats occupations as collections of tasks, and posits that technology substitutes for specific tasks rather than for workers directly. The framework gave rise to the Routine-Biased Technological Change (RBTC) hypothesis, which explained the ``hollowing out'' of the middle class and the U-shaped polarization of the labor market in the late 20th and early 21st centuries \cite{autor2013growth}. Information Technology targeted ``routine'' tasks of two kinds: cognitive routines such as bookkeeping, and manual routines such as assembly-line work. Both could be codified into deterministic algorithms. By contrast, ``non-routine'' tasks that required abstract problem-solving, creativity, or complex physical adaptation were treated as safe harbours for human capital.

\subsection{The LLM Shock: Targeting Non-Routine Cognitive Work}
The emergence of LLMs such as GPT-4 and Llama 3 between 2023 and 2026 has ruptured the RBTC paradigm. Recent work has shown that the primary targets of modern generative AI are precisely the non-routine cognitive tasks long thought immune to automation. Eloundou et al. (2024) used GPT-4 to evaluate the O*NET database and concluded that roughly $80\%$ of the U.S. workforce could have at least $10\%$ of their tasks affected by LLMs. Felten et al. (2023) showed that highly educated white-collar professions, including management analysts, lawyers, and software engineers, exhibit the highest theoretical exposure to AI advances.

Acemoglu (2024), however, cautions that high ``exposure'' does not translate seamlessly into productivity gains or outright substitution. The current wave of optimism tends to ignore the micro-level frictions of deployment, and the literature would benefit from more granular, risk-adjusted measurement frameworks.

\subsection{Risk Aversion, Hallucinations, and Moravec's Paradox}
To understand the friction between technical exposure and actual labor substitution, the discussion needs to integrate insights from behavioral economics, epistemology, and robotics. The translation of theoretical AI exposure into actual substitution is constrained by the epistemological limits of current statistical-fitting architectures \cite{Bubeck2023, DellAcqua2023}. LLMs fail unpredictably in long-tail scenarios, a phenomenon known as the ``Jagged Technological Frontier.'' Deploying these models in high-stakes environments therefore triggers liability asymmetries that traditional task-based models structurally ignore \cite{Agrawal2023, Autor2024}.

The cost of these failures is asymmetrical. Prospect Theory \cite{kahneman1979prospect} highlights human \textit{Loss Aversion}, which explains why management structures resist delegating high-stakes decisions to black-box algorithms. The resilience of physical labor against AI penetration is best explained by \textit{Moravec's Paradox} \cite{moravec1988mind}: high-level reasoning requires modest computation, while low-level sensorimotor skills require enormous, often currently insurmountable, resources. Polanyi's Paradox \cite{polanyi1966tacit} reinforces the point by observing that much of human physical and professional expertise relies on tacit knowledge that cannot easily be textualized for LLM training.

Collectively, this literature suggests a critical void: the necessity of a model that simultaneously measures an AI's cognitive reach while aggressively penalizing its deployment based on real-world commercial and physical risks. This paper fills this void by quantifying both dimensions at the atomic action level.

% =====================================================================
% Chapter 3: Methodology and Data Pipeline
% 这部分重构了整个方法论，从底层数据抓取到多模型打分，再到最核心的 3x3 矩阵抽样验证
% =====================================================================
\section{Methodology and Data Pipeline}

To systematically evaluate the labor market impact of LLMs, this study adopts a bottom-up, task-based methodological framework. We deconstruct macro-occupations into atomic work activities, employ a multi-agent AI ensemble for large-scale capability and risk scoring, and strictly validate the results through a human-in-the-loop stratified sampling approach.

% ---------------------------------------------------------------------
% 3.1 节：数据基座（解释为什么用 DWA 作为最小颗粒度）
% ---------------------------------------------------------------------
\subsection{Data Foundation: Deconstructing Occupations into Atomic Actions}

The foundational dataset for this research is the O*NET (Occupational Information Network) database, version 30.2. Traditional analyses evaluate AI impact at the macro-occupational or task level. A ``task'' such as ``maintain network security,'' however, bundles multiple distinct cognitive and physical actions and creates evaluation ambiguity.

To eliminate this ambiguity, we selected the \textit{Detailed Work Activity (DWA)} as the minimum granular unit of analysis. DWAs represent atomic, indivisible work actions, such as ``Examine crime scenes to obtain evidence,'' stripped of their broader occupational context. By extracting the complete set of 2,087 DWAs, we constructed a standardized, cross-industry action taxonomy. This granular foundation lets our evaluation isolate the underlying nature of the action from the occupational title that hosts it.

% ---------------------------------------------------------------------
% 3.2 节：AI 大规模打分（交待 4 个模型和 2 个维度）
% ---------------------------------------------------------------------
\subsection{AI-Driven Mass Scoring: The Multi-Model Ensemble}

Given the scale of the dataset (2,087 DWAs), relying solely on human expert evaluation is both cost-prohibitive and susceptible to individual subjective fatigue. Therefore, we deployed an ensemble of four state-of-the-art Large Language Models (Qwen, Gemma, Llama, and Mistral) to conduct the primary mass scoring. 

To eliminate the irreproducibility introduced by dynamic updates to closed-source APIs, we constructed a fully localized, open-source LLM scoring matrix. The ensemble runs on a dual NVIDIA RTX 3090 (24GB VRAM) setup and incorporates four representative instructional-tuned frontier models tuned via Reinforcement Learning from Human Feedback \cite{ouyang2022training}: Qwen2.5-32B-Instruct, Gemma-2-27b-it, Meta-Llama-3.1-8B-Instruct, and Mistral-Nemo-Instruct-2407. To balance VRAM consumption against inference precision, we quantized all models uniformly using the Q4\_K\_M GGUF format.

The models were prompted to act as objective capability assessors, evaluating each DWA across two orthogonal dimensions (the explicit zero-shot prompting template, defining the precise boundary conditions for the Tech Level and Risk Score, is detailed in Appendix A):
\begin{itemize}
    \item \textbf{Tech Level (0-3):} The technical feasibility of current AI agents executing the action autonomously.
    \item \textbf{Risk Score (1-5):} The potential business, legal, and safety consequences of an AI failure during execution (ranging from 1 = minor inefficiency, to 4 = severe litigation risk, and 5 = physical injury or fatality).
\end{itemize}
The use of a four-model ensemble mitigates the idiosyncratic biases inherent in any single LLM, providing a robust baseline of algorithmic consensus.

% =========================================================
% 插入图 3：2000+ DWA 动作聚类气泡图 (纯黑字紧凑箭头版 PDF)
% =========================================================
\begin{figure}[htbp]
    \centering
    \includegraphics[width=0.85\linewidth]{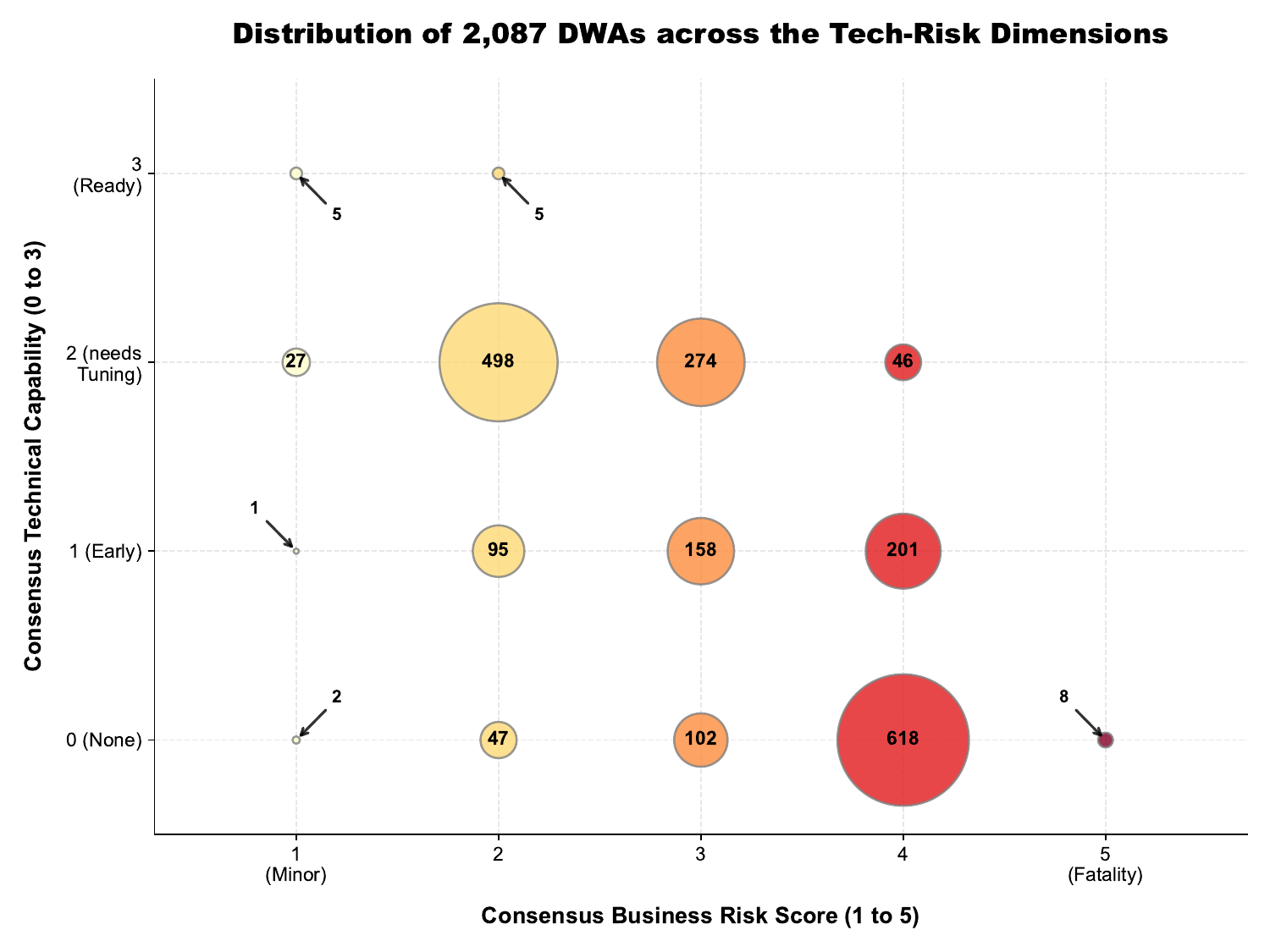}
    \caption{Distribution of 2,087 Detailed Work Activities (DWAs) across the orthogonal dimensions of Technical Capability and Business Risk. Bubble area represents the frequency of atomic actions, with a colour gradient reflecting risk escalation. The dense clustering in medium-to-high risk zones ($R \ge 3$) reflects the friction of real-world AI deployment.}
    \label{fig:dwa_bubble}
\end{figure}

% ---------------------------------------------------------------------
% 3.3 节：人类在回路的 3x3 矩阵抽样验证（极其硬核的数据展示环节）
% 这里将刚刚通过代码计算出的矩阵数据全盘脱出，极其震撼
% ---------------------------------------------------------------------
\subsection{Human-in-the-Loop Validation: A Stratified 3$\times$3 Matrix Analysis}

To validate the reliability of the AI-generated scores, we designed a rigorous Human-in-the-Loop (HITL) validation protocol. Rather than random sampling, we employed a \textit{variance-based stratified sampling} technique. We calculated the scoring variance ($\sigma^2$) among the four AI models for the Risk Score of each DWA and divided a 100-sample dataset into three strata: \textit{Consensus Zone} ($\sigma^2 = 0, n=49$), \textit{Slight Friction Zone} ($\sigma^2 = 0.25, n=17$), and \textit{Severe Divergence Zone} ($\sigma^2 \ge 0.33, n=34$). 

We recruited 31 domain experts for the double-blind evaluation. To study this intersection of algorithms and human society, an emerging discipline conceptualized as \textit{Machine Behaviour} \cite{rahwan2019machine}, we designed the expert panel for ecological validity rather than survey size. A sample of 31 is moderate for general population surveys but is robust for elite, specialized organizational-behavior studies, given the acquisition cost of highly qualified, boundary-spanning practitioners. The panel was geographically diverse (United States, China, South Korea) and professionally heterogeneous, drawing from publicly traded technology conglomerates and specialized business process outsourcing (BPO) firms. We segmented the experts into two macro-cohorts: a \textit{Technology Cohort} ($n=11$, focused on algorithmic boundaries) and a \textit{Risk \& Management Cohort} ($n=20$, covering legal, ethics, HRM, and corporate management). To isolate genuine commercial risk aversion from mere technological ignorance, we ran an ``Epistemic Qualification'' protocol with the Risk \& Management Cohort before the evaluation. Survey work on AI adoption often suffers from a specific endogeneity: respondents' resistance reflects a misunderstanding of current AI capabilities rather than an informed assessment of deployment risks. By pre-calibrating the cohort's understanding of contemporary AI paradigms (AI Agents, Model Context Protocol, platform ecosystems), we controlled for this epistemic deficit.

The statistical alignment on the \textbf{Tech Level} (overall Spearman $\rho = 0.876, p = 8.60 \times 10^{-33}$) serves as a manipulation check: human evaluators accurately grasped the technological frontier. The cognitive gap observed in the \textbf{Risk Score} evaluation (the $+0.35$ inflation) can therefore be isolated cleanly. It is not an artifact of ignorance, but a structural penalty: \textit{Cognitive Risk Asymmetry}, applied by experts who understand the algorithm's capabilities but remain bound by institutional loss aversion and legal accountability.

A marked cognitive gap emerged in the \textbf{Risk Score} evaluation. To dissect this discrepancy, we constructed a $3 \times 3$ analytical matrix (Strata $\times$ Evaluator Cohort) to map the mean perceived risk scores (Table \ref{tab:3x3_matrix}).

\begin{table}[ht]
\centering
\caption{The $3 \times 3$ Risk Score Matrix: Algorithmic Probability vs. Human Perception}
\label{tab:3x3_matrix}
\begin{tabular}{p{4.5cm}cccc}
\toprule
\textbf{Strata (Based on AI Variance)} & \textbf{N} & \textbf{AI Models} & \textbf{Tech Cohort} & \textbf{Risk \& Mgmt Cohort} \\
\midrule
Consensus Zone ($\sigma^2 = 0$) & 49 & 3.80 & 3.62 & 3.65 \\
Slight Friction Zone ($\sigma^2 = 0.25$) & 17 & 3.31 & 3.30 & 3.45 \\
Severe Divergence Zone ($\sigma^2 \ge 0.33$) & 34 & 2.51 & 2.69 & 2.86 \\
\bottomrule
\end{tabular}
\end{table}

The matrix reveals a striking behavioral pattern. In the \textit{Consensus Zone}, dealing with unambiguously extreme cases (either evidently harmless or highly fatal), all cohorts demonstrated strong alignment. However, in the ambiguous \textit{Severe Divergence Zone}, the pure algorithmic probability assessment exhibited significantly lower perceived risk. The Technology Cohort applied a moderate penalty, whereas the Risk \& Management Cohort imposed a severe Cognitive Risk Asymmetry. Because the AI evaluations (derived from mean internal logit probability distributions mapped to integers) and human evaluations (anchored in institutional reality) possess non-equivalent measurement scales, direct arithmetic subtraction of their absolute means is econometrically flawed. 

To assess this cognitive gap, we set aside linear parametric comparisons in favor of an Ordered Logit Model (OLM) and a non-parametric Wilcoxon signed-rank test. The Wilcoxon test confirmed divergence in the matched pairs ($W = 130.5, p < 0.01$). Defining evaluator identity as a dummy variable ($0 = \text{AI}, 1 = \text{Human Expert}$), the OLM shows that human institutional evaluators are structurally more likely to assign higher ordinal risk ratings to the same ambiguous DWAs ($\beta = 0.65, p < 0.001$). The $+0.35$ divergence is not a ``bias'' or a ``panic'' to be corrected. It reflects a fundamental tension: the statistical model produces a cold text-based inference, while the human evaluation carries an ``Institutional Premium.'' From a management-science perspective, this premium is a direct empirical quantification of institutional \textit{Algorithmic Aversion} \cite{dietvorst2015algorithm}. It aligns with the literature on task-dependent algorithm aversion \cite{castelo2019task}: human operators who bear ``skin-in-the-game'' legal liability raise the risk threshold in high-stakes regulatory domains where probabilistic algorithmic failures are met with zero-tolerance penalties. The overall Spearman rank correlations for risk perception corroborate this gradient: Management Experts showed lower alignment with the pure algorithmic baseline ($\rho = 0.526, p < 0.001$), while Technology Experts aligned more closely ($\rho = 0.569, p < 0.001$).

% =========================================================
% 插入图 4：认知偏差 3x3 分组柱状图 (PDF)
% =========================================================
\begin{figure}[htbp]
    \centering
    \includegraphics[width=0.85\linewidth]{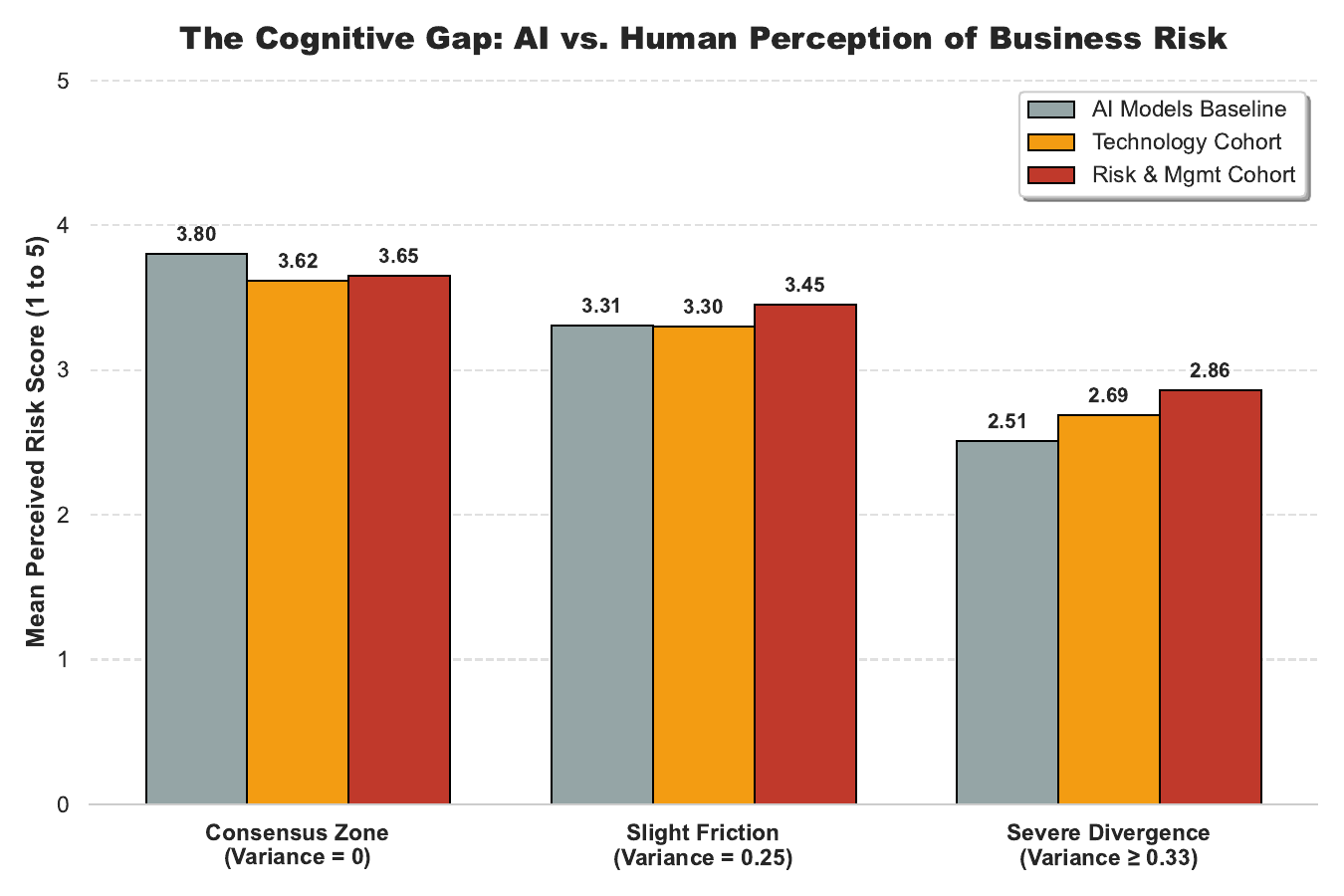}
    \caption{Visualizing the Cognitive Gap in Risk Perception. In the \textit{Consensus Zone}, human experts and AI models align closely. However, in the \textit{Severe Divergence Zone}, human management experts exhibit strong loss aversion, inflating the perceived risk score by +0.35 compared to the objective AI baseline. This asymmetry highlights the friction between purely statistical capabilities and rational institutional risk pricing.}
    \label{fig:cognitive_gap}
\end{figure}

% ---------------------------------------------------------------------
% 3.4 节
% ---------------------------------------------------------------------
\subsection{Finalization of DWA Scores: Non-Linear Risk and the "No Double Penalization" Rationale}

Following the HITL validation, calculating the final operational scores for the 1,987 unvalidated DWAs required aggregating the continuous mean scores generated by the AI ensemble into discrete integers to map onto the Tech-Risk Dual-Factor Model (Section 4.1). For both the technical capability dimension ($T_i$) and the risk dimension ($R_i$), we applied standard nearest-integer rounding (e.g., $\text{round}(3.25) = 3$) to the AI ensemble's arithmetic mean. We deliberately chose \textit{not} to mathematically calibrate the AI-generated risk scores upwards to match the $+0.35$ inflation observed in human management experts (Table 2).

This decision is grounded in the need to avoid a structural ``Double Penalization'' in our model. The epistemic gap between algorithmic outputs and human judgments is wide: the human scale for risk is non-linear and shaped by survival and institutional liability. As behavioral economics and legal theory have shown \cite{kahneman1979prospect}, and as Kleinberg et al. \cite{kleinberg2018human} observed for machines versus human decision thresholds, humans in high-stakes environments systematically apply a disproportionate penalty to catastrophic tail-end risks. The transition from Risk Level 3 (moderate business loss) to Level 4 (legal liability) or Level 5 (physical fatality) triggers a massive ``Institutional Premium.'' Human evaluators, acting as fiduciary guardians, elevate baseline scores in ambiguous scenarios precisely because they must absorb the consequences.

Post-evaluation qualitative interviews with the management cohort completely corroborate this dynamic. Evaluators explicitly described a fundamental cognitive phase shift: while Levels 1 to 3 were perceived as scalable magnitudes of functional friction, the transition to Level 4 (litigation) and Level 5 (fatality) represented an absolute qualitative, institutional boundary. This structural discontinuity invalidates simple linear measurement and highlights the deep rationality underlying the human expert penalty identified by our Ordered Logit Model.

However, from an econometric modeling perspective, injecting this human institutional premium directly into the initial input variable ($R_i$) would conflate the task's textual risk profile with the organization's downstream risk mitigation strategy. The LLM generative ensemble, as a statistical construct operating outside the bounds of human jurisprudence and biological safety, provides an unadulterated baseline of theoretical danger probability.

The commercial resistance stemming from these liabilities is operationalized through the non-linear constraints embedded in the Tech-Risk mapping function (Equation 1). An algorithmic risk baseline of $R=4$ triggers a structural penalty that caps the Automation Index at $AI=0.3$; $R=5$ enforces an absolute veto ($AI=0$). If we were to fold the human institutional premium into the raw AI probability (inflating the $R_i$ input to absorb accountability pressures), and then pass that hybridized score through the matrix's aggressive degradation filter, we would penalize the AI twice for the same risk. By keeping the pure algorithmic probability (the input) separate from the human-designed institutional friction (the mapping formula), we maintain a clean and theoretically defensible foundation for macroeconomic forecasting.

% 4.1
\section{The Occupational Replaceability Model}

\subsection{Construction of the Automation Index via Tech-Risk Dual-Factor Mapping}

In traditional Task-Based Approach evaluations, pioneering studies on the labor-market impact of LLMs, including Eloundou et al. (2024), have focused on the ``exposure'' of specific tasks to technological capabilities. This unidimensional technological perspective overlooks a critical friction in real-world commercial adoption: the cost of error and compliance risk.

Current generative AI paradigms operate primarily on statistical fitting, mapping probability distributions across massive datasets rather than possessing true causal understanding or logical reasoning about the physical world. This probability-based generation produces hallucinations and critical failures in long-tail edge cases. Such failures are acceptable in high-tolerance text-generation settings, but their cost is amplified in core business environments involving financial security, legal compliance, or physical safety. We therefore introduce the ``Risk Score'' as an orthogonal dimension to the ``Tech Level,'' and construct a Tech-Risk Dual-Factor Mapping Matrix that captures actual automation potential rather than theoretical capability.

For any Detailed Work Activity (DWA) defined in the O*NET database, we define its Automation Index ($AI$) as the probability that the activity will be fully automated and stripped of human involvement by LLMs and related autonomous agents within the next 1 to 3 years. Let $T_i \in \{0, 1, 2, 3\}$ denote the consensus technical capability score for the $i$-th DWA, and $R_i \in \{1, 2, 3, 4, 5\}$ denote its corresponding business risk score. The Automation Index is calculated via a piecewise mapping function $f(T_i, R_i)$, formulated as follows:

\begin{equation}
AI(DWA_i) = f(T_i, R_i) = 
\begin{cases} 
0, & \text{if } R_i = 5 \text{ or } T_i = 0 \\
1.0, & \text{if } T_i = 3 \text{ and } R_i \le 2 \\
0.7, & \text{if } (T_i = 3 \text{ and } R_i = 3) \text{ or } (T_i = 2 \text{ and } R_i \le 2) \\
0.5, & \text{if } T_i = 2 \text{ and } R_i = 3 \\
0.3, & \text{if } (T_i \in \{2, 3\} \text{ and } R_i = 4) \text{ or } (T_i = 1 \text{ and } R_i \le 3) \\
0, & \text{if } T_i = 1 \text{ and } R_i = 4 
\end{cases}
\end{equation}

These discrete adoption thresholds constitute a formal Bounding Analysis. Following Agrawal, Gans, and Goldfarb \cite{agrawal2019artificial}, we posit that while artificial intelligence sharply reduces the ``cost of prediction'' and generation, it does not lower the ``cost of judgment'' and may in fact raise it. We frame automation adoption as a profit-maximizing decision under uncertainty. Let an enterprise's marginal benefit of predicting or generating via AI for task $i$ be $MB(T_i)$, and let the expected liability penalty of a hallucination be $E[L|R_i]$. The socially optimal adoption rate $AI^*$ is bounded by the first-order condition where marginal productivity gains equal marginal compliance costs: $MB'(T_i) = \frac{\partial E[L|R_i]}{\partial AI}$. Because legal frameworks governing high-risk environments ($R_i \ge 4$) often enforce \textit{Strict Liability} \cite{shavell1980strict}, asymmetric tail-risk penalties apply, with $E[L|R=4] \gg E[L|R=3]$. The optimal adoption boundary therefore degrades into discrete regulatory regimes. The $0.3$, $0.5$, and $0.7$ thresholds serve as heuristic boundary conditions for these endogenous equilibrium states, where the legal necessity of a ``Human-in-the-Loop'' reviewer offsets the residual liability. We treat these values as open-source parameters for future macroeconomic modeling; subsequent work can recalibrate them against granular corporate actuarial and insurance settlement data. Rather than claiming numerical finality, our matrix models the structural step-function decay of algorithmic utility when it meets strict institutional liability.

% =========================================================
% 插入图 1：技术-风险热力图
% =========================================================
\begin{figure}[htbp]
    \centering
    \includegraphics[width=0.85\linewidth]{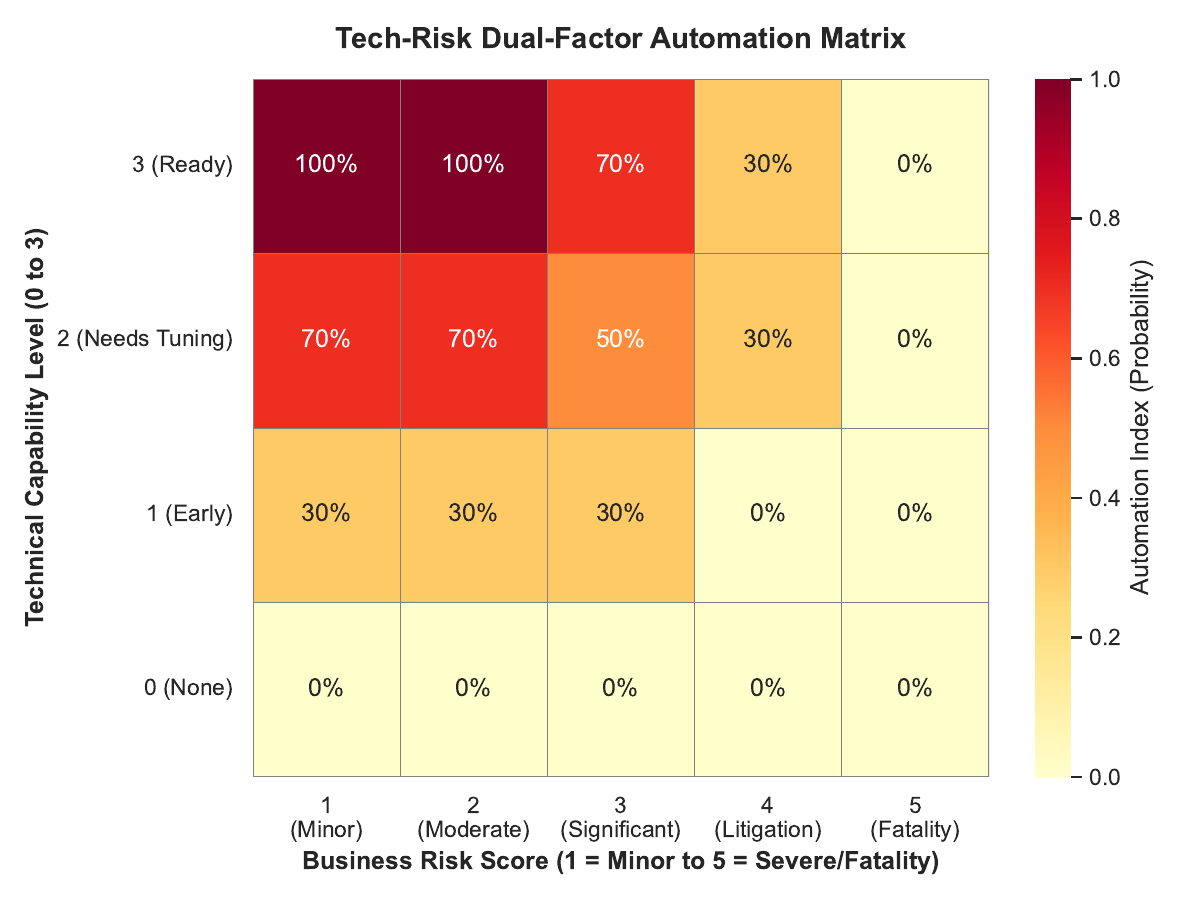} % 如果你存的是png，把后缀改为.png
    \caption{The Tech-Risk Dual-Factor Automation Matrix. The color gradient represents the Automation Index ($AI$), illustrating the non-linear penalization of technical capabilities by business, legal, and safety risks.}
    \label{fig:tech_risk_heatmap}
\end{figure}
% =========================================================

The construction of this dual-factor matrix is not a simple arithmetic decay but a reflection of asymmetric risk tolerance principles deeply rooted in behavioral economics and management decision theory. It is characterized by three core logical pillars:

\begin{itemize}
    \item \textbf{The Veto Power of Severe Risk:} The matrix dictates that when $R_i = 5$ (indicating that task failure could lead to severe physical harm or systemic destruction), the Automation Index is forced to zero ($AI = 0$), regardless of the technical capability ($T_i$). Mathematically, this enforces Moravec's paradox within labor market evaluations. Until the AI paradigm evolves from probabilistic fitting to interpretable logical reasoning with embodied physical causal understanding, core tasks involving unstructured physical interaction cannot be safely or entirely automated.
    
    \item \textbf{Degradation to Co-pilot under Compliance Constraints:} When a task presents significant legal, reputational, or compliance risks ($R_i = 4$), the Automation Index is capped at $0.3$ even when the AI demonstrates maximum technical capability ($T_i = 3$). This threshold captures the ``accountability dilemma'' in commercial deployment. For critical tasks such as drafting binding legal contracts or finalizing financial audits, organizations must degrade the AI to an augmentation tool, that is, a co-pilot. The human worker's role pivots from executor to reviewer and primary liability bearer, which prevents full occupational displacement.
    
    \item \textbf{The Trade-off between Technical Readiness and Marginal Risk:} In zones with manageable risk ($R_i \le 3$), the automation probability exhibits a smooth gradient descent. For instance, when a technology is in its theoretical infancy ($T_i = 1$) but carries moderate business risk, the Return on Investment (ROI) for technological adoption falls below the cost of human labor, yielding a near-zero replacement rate. Conversely, frictionless substitution ($AI = 1.0$) only emerges when the technology is fully mature ($T_i = 3$) and the associated risk is negligible ($R_i \le 2$).
\end{itemize}

Through this matrix mapping, the proposed model effectively filters the objective technical capabilities of foundational AI models through the lens of commercial risk tolerance. It is critical to explicitly define the resulting Automation Index ($AI$) not as a deterministic ``Absolute Substitution Forecast,'' but as a calibrated ``Relative Vulnerability Index.'' The discrete degradation boundary conditions (0.3, 0.5, 0.7) represent structurally grounded inflection points marking the burden of liability; however, they remain heuristic parameters pending precise empirical calibration against granular corporate actuarial and insurance settlement data. Nonetheless, they provide a robust, granular data foundation for the subsequent upward aggregation mapping the relative exposure hierarchy of the occupational landscape.

% 4.2
\subsection{Upward Aggregation: From Detailed Work Activities to Occupational Replaceability}

Having established the Automation Index ($AI$) at the granular DWA level through the Tech-Risk dual-factor matrix, the next step is to map these micro-level probabilities back to the macro-economic landscape. The O*NET database provides a hierarchical structure that links occupations to their constituent tasks, and tasks to their underlying DWAs.

We formalize this hierarchical mapping as a set of bipartite graphs. Let $\mathcal{O}$ represent the set of all occupations, $\mathcal{T}$ represent the set of all tasks, and $\mathcal{D}$ represent the set of all DWAs. The mapping from tasks to DWAs is defined by the relation $M_{TD} \subseteq \mathcal{T} \times \mathcal{D}$, and the mapping from occupations to tasks is defined by $M_{OT} \subseteq \mathcal{O} \times \mathcal{T}$.

\textbf{Step 1: Task-Level Aggregation.} 
A specific task $t_j \in \mathcal{T}$ typically comprises a subset of specialized actions, denoted as $\mathcal{D}(t_j) = \{d \in \mathcal{D} \mid (t_j, d) \in M_{TD}\}$. Tasks in high-stakes environments are not additive collections of independent actions; they exhibit strong complementarities, so a single critical failure can collapse the value of the entire process. Following Kremer's O-Ring Theory of Economic Development \cite{kremer1993o}, we abandon the unweighted arithmetic mean, which suffers from linear dilution. Consider a task comprising five DWAs: four algorithmically trivial text-generation actions with $AI=1.0$, and one physical or legally fatal action with $AI=0$. The linear mean yields a misleading $AI=0.8$, when in reality the inability to close the fatal safety loop forces the entire task into a Human-in-the-Loop co-pilot configuration. We therefore formulate the task-level automation index as a bottleneck Leontief-style aggregation:

\begin{equation}
AI(t_j) = \min_{d \in \mathcal{D}(t_j)} AI(d)
\end{equation}

By extracting the lowest constituent DWA substitution rate, this function mathematically enforces the ``veto power of severe risk'' (Section 4.1) at the macroscopic task level. Tasks with a high $AI(t_j)$ strictly indicate that \textit{all} of their operational steps can be reliably delegated to AI agents without encountering an insurmountable physical or commercial bottleneck.

\textbf{Step 2: Occupation-Level Aggregation (Importance-Weighted).} 
An occupation $o_k \in \mathcal{O}$ is defined as a collection of distinct tasks, denoted as $\mathcal{T}(o_k) = \{t \in \mathcal{T} \mid (o_k, t) \in M_{OT}\}$. Unlike simpler models that treat tasks as equally important to an occupation, we use the ``Task Importance'' metric provided by O*NET to construct an importance-weighted aggregation.

Let $I(t)$ represent the normalized importance score of task $t$ within occupation $o_k$. The relative weight $w_t$ of a specific task is determined by its importance relative to the cumulative importance of all tasks defining that occupation:
\begin{equation}
w_t = \frac{I(t)}{\sum_{t' \in \mathcal{T}(o_k)} I(t')}
\end{equation}

The final Occupational Automation Index ($OAI$) for occupation $o_k$ is thus computed as the importance-weighted sum of its constituent task automation probabilities:
\begin{equation}
OAI(o_k) = \sum_{t \in \mathcal{T}(o_k)} w_t \cdot AI(t)
\end{equation}

This weighted approach ensures that the $OAI$ reflects disruption to an occupation's core workflows rather than to its tangential duties. A high $OAI(o_k)$ indicates that the tasks most central to value creation in the occupation are highly susceptible to automation. Economically, this implies that workers in this occupation will reallocate their cognitive bandwidth toward the remaining lower-weight but high-friction tasks, transforming the nature of the occupation.

% 4.3
\subsection{Industry-wide Replaceability Ranking and Trait Analysis}

Applying the importance-weighted Occupational Automation Index ($OAI$) model to the O*NET database yields a vulnerability ranking of 923 distinct occupations. The empirical distribution of the $OAI$ scores reveals a paradigm shift in the automation landscape and extends the Routine-Biased Technological Change (RBTC) hypothesis \cite{Autor2003, Acemoglu2018} to the substitution trajectories of non-routine cognitive tasks.

Our risk-adjusted findings diverge from early, purely technical automation benchmarks. Frey and Osborne \cite{frey2017future} predicted that 47\% of U.S. employment faced categorical obsolescence, and Webb \cite{webb2020impact} identified aggressive exposure among high-skill cognitive roles; our Dual-Factor model shows that commercial viability sharply compresses this vulnerability space. We corroborate Webb's \cite{webb2020impact} thesis that non-routine cognitive work is the primary target of modern neural architectures, but our incorporation of the Risk Score reveals that the absolute majority of occupations remain insulated by compliance friction.

% =========================================================
% 插入图 2：宏观职业分布 KDE 图 (PDF版)
% =========================================================
\begin{figure}[htbp]
    \centering
    \includegraphics[width=0.9\linewidth]{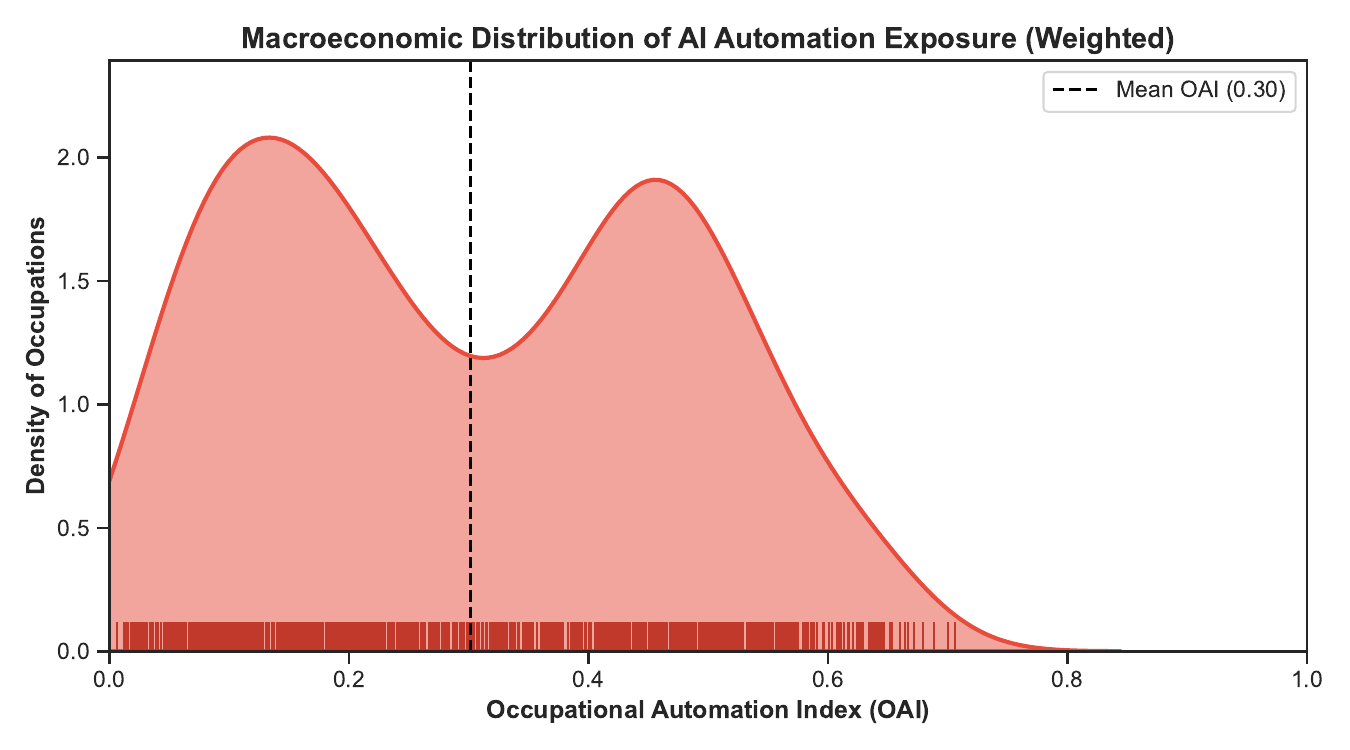}
    \caption{Macroeconomic Labor Market Vulnerability Distribution of the Occupational Automation Index (OAI). The density plot highlights that full replacement is an illusion for the majority of occupations, with exposure concentrated in specific cognitive domains rather than a uniform market-wide displacement.}
    \label{fig:oai_distribution}
\end{figure}

% =========================================================
% 插入 Table 3：全美劳动力市场影响汇总统计表
% =========================================================
\begin{table}[htbp]
\centering
\caption{Summary Statistics of AI Automation Exposure Across the U.S. Labor Market}
\label{tab:summary_stats}
\small
\begin{tabular}{p{4.5cm} c c p{4.5cm}}
\toprule
\textbf{Exposure Category (OAI Range)} & \textbf{Occupations Count} & \textbf{\% of Total} & \textbf{Dominant Occupational Traits} \\
\midrule
\textbf{High Exposure} ($OAI \ge 0.60$) & 41 & 4.4\% & Purely Cognitive, Symbolic Manipulation \\
\textbf{Medium Exposure} ($0.30 \le OAI < 0.60$) & 408 & 44.2\% & Mixed Routine and Non-Routine, Moderate Risk \\
\textbf{Low Exposure / High Resilience} ($OAI < 0.30$) & 474 & 51.4\% & Physical Embodiment, High-Liability \\
\midrule
\textbf{Total Analyzed} & \textbf{923} & \textbf{100.0\%} & -- \\
\bottomrule
\end{tabular}
\vspace{1ex}
\raggedright
\textit{Note:} The distribution shows considerable structural resilience. Once core-task importance weights are integrated, complete vulnerability (High Exposure) is constrained to a small minority of the labor market ($4.4\%$). The absolute majority ($>50\%$) remains heavily insulated by commercial and physical risk frictions.
\end{table}

Historically, technological disruptions hollowed out middle-skill, routine jobs \cite{autor2013growth}, creating a U-shaped labor polarization that preserved low-skill manual labor and high-skill cognitive professions. However, our dual-factor analysis indicates that LLMs and autonomous agents exhibit a distinctly different substitution trajectory: one that is highly aggressive toward the high-skill non-routine cognitive peak, but strictly bounded by physical and risk constraints.

\textbf{The Vulnerability of Symbolic Manipulation (The Top Tier)}
Analysis of the highest-ranking occupations reveals an unprecedented exposure of advanced cognitive and creative roles. Leading the vulnerability index are Data Scientists ($OAI = 0.7062$), Editors ($OAI = 0.7005$), Mathematicians ($OAI = 0.6882$), and Technical Writers ($OAI = 0.6718$). 

% =========================================================
% 职业大表
% =========================================================
\begin{table*}[htbp]
\centering
\caption{Top 15 and Bottom 15 Occupations by Occupational Automation Index (OAI)}
\label{tab:oai_ranking}

\resizebox{\textwidth}{!}{%
\begin{tabular}{llc | llc}
\toprule
\multicolumn{3}{c|}{\textbf{Top 15: Highest Automation Exposure}} & \multicolumn{3}{c}{\textbf{Bottom 15: Highest Resilience (Lowest Exposure)}} \\
\cmidrule(r){1-3} \cmidrule(l){4-6}
\textbf{SOC Code} & \textbf{Occupation Title} & \textbf{OAI} & \textbf{SOC Code} & \textbf{Occupation Title} & \textbf{OAI} \\
\midrule
15-2051.00 & Data Scientists & 0.7062 & 47-4051.00 & Highway Maintenance Workers & 0.0150 \\
27-3041.00 & Editors & 0.7005 & 37-2011.00 & Janitors and Cleaners... & 0.0145 \\
15-2021.00 & Mathematicians & 0.6882 & 49-9051.00 & Electrical Power-Line Installers... & 0.0140 \\
43-9022.00 & Word Processors and Typists & 0.6794 & 47-4061.00 & Rail-Track Laying and Maintenance... & 0.0136 \\
27-3042.00 & Technical Writers & 0.6718 & 47-5071.00 & Roustabouts, Oil and Gas & 0.0120 \\
27-1022.00 & Fashion Designers & 0.6667 & 47-2071.00 & Paving, Surfacing, and Tamping... & 0.0064 \\
15-2051.01 & Business Intelligence Analysts & 0.6642 & 47-3014.00 & Helpers--Painters, Paperhangers... & 0.0000 \\
15-1255.01 & Video Game Designers & 0.6601 & 47-2051.00 & Cement Masons and Concrete Finishers & 0.0000 \\
17-3011.00 & Architectural and Civil Drafters & 0.6537 & 47-2043.00 & Floor Sanders and Finishers & 0.0000 \\
43-9021.00 & Data Entry Keyers & 0.6516 & 47-2072.00 & Pile Driver Operators & 0.0000 \\
19-3093.00 & Historians & 0.6470 & 47-5043.00 & Roof Bolters, Mining & 0.0000 \\
43-9081.00 & Proofreaders and Copy Markers & 0.6461 & 35-9021.00 & Dishwashers & 0.0000 \\
27-3043.05 & Poets, Lyricists and Creative Writers & 0.6439 & 51-3023.00 & Slaughterers and Meat Packers & 0.0000 \\
43-4171.00 & Receptionists and Info Clerks & 0.6426 & 29-1022.00 & Oral and Maxillofacial Surgeons & 0.0000 \\
15-1243.00 & Database Architects & 0.6424 & 29-1024.00 & Prosthodontists & 0.0000 \\
\bottomrule
\end{tabular}%
} % 结束 resizebox
\vspace{1ex}
\raggedright
\small
\textit{Note:} OAI represents the probability of task restructuring. Roles with $OAI \approx 0.70$ face severe cognitive substitution, whereas roles with $OAI < 0.02$ are completely insulated by Moravec's paradox and liability premiums.
\end{table*}
% =========================================================

These occupations share a fundamental trait: their core workflows sit entirely within the digital domain and rely heavily on symbolic manipulation, information processing, and pattern recognition. Because the consequences of failure (a buggy line of code, a syntactical error in a manuscript) typically manifest as correctable business inefficiencies ($R \le 3$) rather than catastrophic physical harm, these tasks bypass the high-risk penalty in our matrix. Cognitive bandwidth, long treated as the ultimate sanctuary of human labor, is highly susceptible to the statistical-fitting and semantic-generation capabilities of modern LLMs.

\textbf{Polanyi's and Moravec's Moats (The Bottom Tier)}
Conversely, the occupations demonstrating the highest resilience to AI substitution are anchored in unstructured physical environments and human-centric care. The bottom of the index is occupied primarily by unpredictable physical labor and high-stakes caregiving, exemplified by occupations such as Roofers ($OAI = 0.0199$) and Home Health Aides ($OAI = 0.0226$). 

Rather than treating the extreme resilience of manual trades such as roofing and stonemasonry and high-stakes healthcare as a novel empirical discovery, our dual-factor matrix functions as a vulnerability distribution that operationalizes \textit{Moravec's Paradox} \cite{moravec1988mind} and \textit{Polanyi's Paradox} \cite{polanyi1966tacit}. By hardcoding absolute veto thresholds ($R=5$) and severe degradation parameters ($R=4$) into our mapping function for unstructured physical interaction and high-liability environments, our model shows how theoretical AI exposure is compressed by real-world physical and legal frictions.

The resulting distribution does not merely reflect raw technical capability; it simulates a risk-adjusted market equilibrium. In this simulated environment, high-level symbolic manipulation is heavily targeted because of its low commercial friction. Low-level sensorimotor skills, by contrast, remain structurally insulated. The insulation has an empirical basis: despite AI's mastery of complex mathematics, traversing an uneven roof or shaping an irregular stone involves long-tail physical variables that current generative models cannot reliably process. This structural resilience is reinforced epistemologically by \textit{Polanyi's Paradox} \cite{polanyi1966tacit}, captured by the axiom that ``we can know more than we can tell.'' Many physical and caregiving actions rely on tacit knowledge and intuitive somatic feedback that resists codification into the text datasets used for LLM training. The intersection of these epistemological limits with the demand for strict human accountability substantiates a \textit{Cognitive Risk Asymmetry} that functions as a moat against full automation in the contemporary labor market.

\textbf{The Cognitive Risk Asymmetry in Healthcare and Infrastructure}
A secondary, yet equally critical, trait of the bottom-tier occupations is the overwhelming presence of severe liability constraints. Roles such as Surgical Assistants and Physical Medicine Physicians ($OAI = 0.0200$) technically involve cognitive diagnosis that an AI could theoretically perform. However, their physical interaction with patients triggers the absolute veto threshold ($R=5$) in our dual-factor model. In these domains, the legal necessity of human accountability and the ethical intolerance for ``probabilistic hallucinations'' construct an impenetrable barrier to full automation, echoing the false hope of current explainable AI approaches in high-stakes healthcare \cite{ghassemi2021false}. Consequently, we observe a substantial `Cognitive Risk Asymmetry'' in the labor market: job security in the AI era is no longer solely dictated by the cognitive complexity of the task, but increasingly by the magnitude of real-world risk associated with its execution.

% 4.4
\subsection{Robustness Check: Sensitivity of the Dual-Factor Matrix}

To ensure that our macro-level conclusions are not merely artifacts of the specific probability thresholds defined in the Tech-Risk Matrix (Section 4.1), we conduct a rigorous robustness check. The fundamental critique of any rule-based mapping function is its sensitivity to parameter adjustments: would the vulnerability hierarchy collapse if commercial risk tolerance radically shifted?

To address this, we defined three distinct macroeconomic scenarios (parallel universes of the labor market) and recalculated the $OAI$ for all 923 occupations:
\begin{itemize}
\item \textbf{Baseline Scenario:} The primary mapping logic utilized in this study, representing the current institutional friction.
\item \textbf{Aggressive Scenario (High Risk Tolerance):} Simulates a hyper-capitalist environment where businesses are willing to force AI adoption despite elevated compliance risks. In this scenario, high-capability AI ($T=3$) achieves full substitution ($AI=1.0$) even at moderate risk levels ($R=3$), and retains $70\%$ substitution even at severe risk ($R=4$).
\item \textbf{Conservative Scenario (Strict Regulation):} Simulates a highly restrictive legal environment. Any task involving significant commercial risk ($R \ge 4$) triggers an absolute ban on AI autonomy ($AI=0$), restricting the model solely to low-risk, deterministic domains.
\end{itemize}

Applying these mapping constraints yields three separate occupational rankings. The analysis shows that while the absolute values (cardinality) of the $OAI$ shift predictably (rising across the board in the aggressive scenario and falling in the conservative one), the \textit{relative structural hierarchy} (ordinality) remains virtually indestructible.

Spearman's rank correlation analysis shows strong consistency across the models. The correlation between the Baseline and Aggressive scenarios is very high ($\rho = 0.9919, p < 0.001$), and remains robust against the Conservative scenario ($\rho = 0.9846, p < 0.001$).

This empirical evidence definitively solidifies our core thesis: while the precise heuristic parameters utilized across the Tech-Risk Matrix currently prevent the calculation of absolute job displacement totals, the underlying vulnerability hierarchy driving the index is not a fragile mathematical construct. The ordinal dominance of the "Cognitive Risk Asymmetry" over pure cognitive capability is a fundamental, invariant structural feature of the impending AI-driven labor market transition.

% 5
\section{Discussion}

The empirical findings of this study, grounded in the Tech-Risk Dual-Factor Model, provide a nuanced departure from the prevailing narrative of imminent, widespread technological unemployment. By disaggregating occupations into DWAs and juxtaposing algorithmic capabilities against commercial risk constraints, several critical insights regarding the future of human-AI labor dynamics emerge.

\subsection{The Cognitive Gap in Risk Perception: Pure Algorithmic Probability vs. Human Institutional Premium}

One of the most striking dynamics isolated during the human-in-the-loop validation phase (Section 3.3) was the structural epistemological divergence in risk perception between the LLM ensemble and human management experts. AI models and experts achieved near-perfect alignment on pure technical boundaries (Spearman $\rho = 0.828, p < 0.001$), but this consensus fractured when business and compliance risks entered the picture. The mean baseline generated by the AI model registered significantly lower than the final assessments rendered by human operators.

We read this divergence not as irrational bias on the part of human evaluators, but as the expression of an ``Institutional Premium.'' From a behavioral standpoint \cite{kahneman1979prospect}, terms such as ``litigation,'' ``safety hazard,'' and ``reputational death'' trigger non-linear asymmetric risk pricing in humans charged with protecting an organization's longevity. From an epistemic standpoint, current LLM paradigms lack the mechanics of consequence internalization, what Taleb calls ``skin in the game'' \cite{taleb2018skin}. As disembodied statistical matrices processing tokens in a vacuum, algorithms cannot be legally sanctioned, financially bankrupted, or physically injured. The $+0.35$ inflation registered by the human cohort is a rational defensive augmentation needed to bridge the gap between statistical probability and the absolute accountability demanded by human institutional frameworks. The point identifies a structural roadblock for current AGI commercialization: until models can compute and mirror holistic compliance friction, human fiduciary agents will remain indispensable as the final arbiters of societal risk.

\subsection{From Substitution to Augmentation: The Human-in-the-Loop Imperative}

Our occupational aggregation shows that even the most highly exposed professions, such as Data Scientists ($OAI = 0.7062$) and Editors ($OAI = 0.6897$), do not reach a complete substitution threshold. This ceiling is governed by the risk penalty embedded in our matrix.

The immediate micro-level impact of LLMs is therefore not mass occupational extinction, but an aggressive intra-occupational task reallocation. As AI absorbs low-to-medium risk execution tasks such as generating boilerplate code and drafting initial analytical reports, the cognitive bandwidth of human workers will be pushed toward the ``high-risk, high-liability'' tail of the task distribution. The nature of white-collar work is poised to transition from \textit{creation} to \textit{curation and auditing}. This aligns with the Susskind and Susskind \cite{susskind2015future} framework on the dismantling of traditional professions, in which highly specialized work is decomposed into routine data processing allocated to algorithms and high-stakes auditing reserved for human experts. This captures the \textit{automation--augmentation paradox} proposed by Raisch and Krakowski \cite{raisch2021artificial}: while the structural intent of AI deployment is complete automation through substitution, the commercial and legal risks force a retreat into augmentation architectures. In this new paradigm, the \textit{Human-in-the-Loop} (HITL) system transitions from a temporary safety measure into a permanent, legally mandated feature of the labor market.

\subsection{Beyond RBTC: The Emergence of the Compliance Premium}

The RBTC hypothesis explained the historical hollowing-out of middle-income jobs by modeling wage polarization across decades. Our $OAI$ metric, by contrast, establishes a static snapshot of technological exposure for the year 2026. Mapping this vulnerability index against the RBTC framework nonetheless uncovers a clear dynamic implication for future wage structures.

As generative AI targets advanced non-routine cognitive work, traditionally the peak of the income distribution, the primary moat protecting human labor ceases to be ``cognitive complexity.'' Extrapolating from our 2026 static cross-sectional exposure data, we propose a hypothesis about long-term wage restructuring: the emergence of a \textbf{Compliance Premium}. Rather than a U-shaped wage polarization driven by the routine vs.~non-routine dichotomy, this hypothesis predicts that the future equilibrium labor market will reallocate wealth toward positions that carry intense regulatory liability and moral hazard. The wage premium may decouple from pure intellectual execution and tether itself to the human capacity to absorb institutional risk. This transformation echoes historical periods in which technology boosted employment through new task creation \cite{bessen2019automation} while catalyzing the rise of alternative work arrangements tailored to specialized risk-bearing activities \cite{katz2019rise}. If the hypothesis holds under dynamic modeling, it would illustrate the macroscopic \textit{Reinstatement Effect} framework established by Acemoglu and Restrepo \cite{acemoglu2019automation}. AI technically displaces human labor from pure cognitive execution, but the demand for legal accountability and subjective moral judgment is hypothesized to \textit{reinstate} human oversight into a new ecosystem of compliance and risk management.

\subsection{Methodological Bounds and Requirements for Equilibrium Modeling}
We need to establish the empirical boundaries of the Occupational Automation Index (\textit{OAI}). The $OAI$ functions strictly as a \textit{static cross-sectional diagnostic of relative systemic vulnerability}, representing a pre-equilibrium technological shock vector in 2026. Direct extrapolation from high $OAI$ values to absolute predictions of macroeconomic unemployment or final equilibrium wage collapse therefore constitutes an econometric overclaim.

Transforming these localized task-level exposure indices into robust macroeconomic outcome forecasts strictly mandates integrating the OAI into formalized Computable General Equilibrium (CGE) models. Extrapolating to the true systemic impact on the labor market necessitates the modeling of complex dynamic constraints, including endogenous capital-labor substitution elasticity (how quickly firms can actually afford to swap humans for AI server capacity), cross-industry labor re-equilibration dynamics (where displaced workers migrate), and shifting product demand elasticities (how lowered prices from AI efficiency spike demand and subsequently drive rehiring, i.e., the productivity effect). Consequently, the Compliance Premium remains a theoretically bounded macroeconomic hypothesis, serving as the necessary impetus for future targeted dynamic modeling.

\subsection{Macroeconomic and Educational Implications: The Case for Strategic Stratification}

The structural inversion of occupational vulnerability highlighted in Section 4.3 presents a severe challenge to contemporary educational policies. For the past two decades, global education systems have aggressively promoted ``universal coding'' and standardized STEM literacy, operating on the assumption that routine symbolic manipulation represents the safest harbor in the future economy. However, our dual-factor model demonstrates that middle-tier cognitive tasks are precisely the most aggressively targeted by generative AI. Consequently, mass-producing entry-level programmers or routine data processors poses a severe risk of structural unemployment.

To mitigate this, educational paradigms must shift from a homogenized curriculum toward \textit{Strategic Stratification} and aptitude-based tracking (echoing the pedagogical philosophy of differentiated instruction). This bifurcation strategy necessitates two distinct educational trajectories:

\textbf{1. Elite Computational Thinking for System Architects:} 
Abandoning computer science education would be a mistake. For top-tier analytical talent, the pedagogical focus must pivot from teaching ``language syntax'' to cultivating high-level \textit{Computational Thinking}. This means training cognitive elites to architect complex logical workflows, design algorithmic systems, and orchestrate multiple AI agents to solve multi-step real-world problems. These individuals will serve as the system designers and final decision-makers who set the boundaries within which AI operates.

\textbf{2. The Renaissance of Embodied and Risk-Managing Professions:} 
For the broader workforce, education must re-center on meta-skills with high friction against AI substitution. Our data shows that the lowest exposure scores are concentrated in roles requiring either physical unpredictability or extreme legal and ethical accountability, which corroborates Deming's \cite{deming2017growing} hypothesis on the secular premium on social and collaborative skills. Vocational tracking should therefore be destigmatized and elevated. We need a new generation of workers specializing in \textit{Advanced Embodied Trades}, the modernization of physically complex blue-collar professions such as advanced infrastructure maintenance and specialized healthcare, as well as roles demanding deep interpersonal empathy, moral reasoning under ambiguity, and complex risk-management judgment. Aligning educational tracking with the absolute comparative advantages of human biology and legal accountability lets society construct a labor market that complements rather than competes with artificial intelligence.

% 6
\section{Conclusion and Future Outlook}

\subsection{Concluding Remarks}

As Large Language Models rapidly transition from experimental laboratories to real-world commercial deployment, accurately predicting their labor market impact requires moving beyond unidimensional evaluations of algorithmic capability. This study introduces the Tech-Risk Dual-Factor Model, applying a bottom-up, task-based approach to deconstruct 2,087 Detailed Work Activities (DWAs) and map them to 923 occupations in the O*NET database.

Through a multi-national Human-in-the-Loop validation protocol and stratified sampling, we demonstrated a cognitive gap between algorithmic probability and human commercial risk perception. By factoring in the ``Cognitive Risk Asymmetry'' driven by legal accountability, safety constraints, and human loss aversion, our findings challenge the Routine-Biased Technological Change (RBTC) hypothesis. The contemporary labor-market moat is no longer defined by cognitive complexity, but by physical friction and liability. Symbolic manipulation and non-routine cognitive professions such as Data Scientists and Editors therefore face unprecedented exposure, while embodied physical trades and high-stakes healthcare roles remain structurally insulated. The immediate macroeconomic future is not mass occupational extinction, but an aggressive transition toward a legally mandated \textit{Human-in-the-Loop} paradigm, in which the core value of human capital shifts from execution to auditing and risk management.

\subsection{Limitations and the Horizon of ``Logical AI''}

While the Tech-Risk Dual-Factor Model provides a robust framework for the current technological landscape, we acknowledge several critical limitations that define the boundaries of our empirical findings.

First, the primary limitation lies in the extreme temporal volatility of the independent variable ($Tech\_Level$). The velocity of advancement in generative AI models and multi-agent systems is unprecedented. The technological frontier is non-stationary; thus, the Occupational Automation Indices ($OAI$) calculated in this study represent a definitive snapshot of the 2026 AI landscape. Given the exponential rate of algorithmic optimization, we postulate that the $Tech\_Level$ baseline will necessitate rigorous recalibration within a compressed 6- to 12-month horizon to maintain predictive validity. 

Second, we acknowledge the constraint of our human expert sample size ($N=31$). The panel was deliberately specialized and elite, which was necessary to isolate genuine institutional premium from mere technological ignorance, but the limited $N$ restricts econometric generalizability across geographic and regulatory ecosystems. Future research should prioritize large-scale global surveys of corporate executives and risk officers to replicate and further calibrate the $+0.35$ ``Institutional Premium'' identified in our behavioral analysis.

Third, reliance on the O*NET taxonomy introduces a temporal lag. O*NET characterizes the historical and present anatomy of labor, so our data is structurally backward-looking. The $OAI$ index isolates the risk of traditional workflows being deconstructed, but it cannot capture the simultaneous creation of novel AI-complementary tasks such as Prompt Engineering, AI Auditing, and LLM Orchestration. Our findings therefore quantify the vulnerability of the existing task matrix rather than offering a holistic forecast of a future labor market that will inevitably integrate new tasks.

Fourth, our projection models the impact of an AI paradigm based on \textit{Statistical Fitting}: systems that map high-dimensional probability distributions but lack deterministic causal reasoning. The absolute resilience of physical and high-liability tasks ($R \ge 4$) in our matrix rests entirely on this epistemological limitation.

As Pearl and Mackenzie \cite{pearl2018book} argue, current deep-learning architectures remain trapped on the lowest rung of the ``Ladder of Causation'' (association and observation); they are computationally incapable of performing true interventions or contemplating counterfactuals. This epistemological deficit is the root cause of their inability to operate autonomously in high-risk physical environments ($R \ge 4$). The anticipated paradigm shift from purely Statistical AI to \textit{Logical AI} \cite{marcus2020next}, that is, Artificial General Intelligence equipped with robust neurosymbolic causal reasoning, system-2 thinking, and embodied integration, will produce a structural shock to the labor market. As Marcus \cite{marcus2020next} argues, moving beyond brittle statistical correlations toward robust causal intelligence is required to navigate the open-ended physical world. When autonomous agents cross the rubicon of Moravec's Paradox by demonstrating absolute reliability in physical interactions and deterministic logic in high-stakes environments, the risk penalties currently insulating bottom-tier occupations will collapse. We hypothesize that the arrival of mature Logical AI will trigger a structural inversion of our current vulnerability rankings, a macroeconomic phenomenon we intend to quantify in subsequent research.

\bibliographystyle{plain}  
\bibliography{references} 

\clearpage
\appendix
\section{AI Ensemble System Prompt}
\label{app:ai_prompt}

The data generation process (DGP) for the technical capabilities and risk metrics relied on a highly constrained zero-shot prompt. The system prompt used across the LLM ensemble is documented below:

\begin{small}
\begin{verbatim}
You are a top-tier assessment expert at the intersection of
labor economics and artificial intelligence. Your task is to
evaluate the given [Detailed Work Activity (DWA)] and score it
across two dimensions: Technical Implementation Path
(tech_level) and Failure Risk Penalty (risk_score).

[Dimension 1: Technical Implementation Path (tech_level)]
Level 3: Native LLM Replacement. Pure text and data processing;
        current LLMs can complete it without external tools.
Level 2: Agent and MCP Integration. The model requires specific
        plugins (such as web search or file reading) to complete
        it fully automatically.
Level 1: System Integration. Technically feasible, but requires
        IT departments to develop APIs to connect legacy
        systems or hardware.
Level 0: Human-in-the-loop Required. Involves complex physical
        world interaction, highly nuanced emotional support, or
        critical moral and legal final decisions. Current AI
        cannot close the loop independently.

[Dimension 2: Failure Risk Penalty (risk_score)]
1: No risk (drafting a document with typos, easily fixable).
2: Minor business impact (sending an incorrect internal email).
3: Moderate loss (losing a single client or causing minor
   financial loss).
4: Severe loss (facing legal action, severe reputation crisis,
   or a major safety incident).
5: Fatal impact (endangering human life, license revocation,
   or company bankruptcy).

[Output Requirements]
You MUST ONLY return a valid JSON object. Do not output any
Markdown formatting, and do not include any conversational
filler. The format must be exactly as follows:
{"tech_level": 2, "risk_score": 3,
 "reasoning": "A brief explanation of why."}
\end{verbatim}
\end{small}

\end{document}